\documentclass[12pt]{article}
\usepackage{epsfig}
\begin{document}

{\large \sf

{\sf
\title{
{\normalsize
\begin{flushright}
CU-TP1137
\end{flushright}}
Convergent Iterative Solutions for a Sombrero-Shaped Potential in
Any Space Dimension and Arbitrary Angular Momentum\thanks{Work
supported in part by the U.S. Department of Energy and the
National Natural Science Foundation of China}}

\author{
R. Friedberg$^{1}$, T. D. Lee$^{1,2}$ and W. Q. Zhao$^{2,3}$\\
{\small \it 1. Physics Department, Columbia University, New York, NY 10027, USA}\\
{\small \it 2. China Center of Advanced Science and Technology (CCAST)}\\
{\small \it         (World Lab.), P.O. Box 8730, Beijing 100080, China}\\
{\small \it 3. Institute of High Energy Physics, Chinese Academy
of Sciences}\\
{\small \it Beijing 100039, China}  } \maketitle
\date{}

\begin{abstract}
We present an explicit convergent iterative solution for the
lowest energy state of the Schroedinger equation with an
$N$-dimensional radial potential $V=\frac{g^2}{2}(r^2-1)^2$ and an
angular momentum $l$. For $g$ large, the rate of convergence is
similar to a power series in $g^{-1}$.
\end{abstract}

\vspace{1cm}

{\normalsize ~~~~PACS{:~~11.10.Ef,~~03.65.Ge}}

\newpage

{\large \sf

\section*{\bf 1. Introduction}
\setcounter{section}{1} \setcounter{equation}{0}

The problem of a non-relativistic particle moving in an
$N$-dimensional Sombrero-shaped potential provides a prototype
example of the spontaneous symmetry breaking mechanism. Yet, even
when $N=1$, it is difficult to solve the corresponding
Schroedinger equation with a quartic potential[1-10]. In this
paper, we shall give explicit convergent iterative solutions for
the Schroedinger equation
$$
H\Psi=(-\frac{1}{2}{\bf \nabla}^2 +V)\Psi=E\Psi \eqno(1.1)
$$
in $N$-dimension and with angular momentum $l$. Let $q$ be the
Cartesian coordinates
$$
q =(q_1,~q_2,~\cdots,~q_N) \eqno(1.2)
$$
with $N>1$ and ${\bf \nabla}^2$ the Laplacian
$$
{\bf \nabla}^2 = \sum_{i=1}^{N}\frac{\partial^2}{\partial q_i^2}.
\eqno(1.3)
$$
The potential is
$$
V = \frac{g^2}{2} (r^2-1)^2,\eqno(1.4)
$$
where
$$
r^2 = \sum_{i=1}^{N} q_i^2.\eqno(1.5)
$$
To illustrate our approach, it may be useful to consider first the
groundstate ($s$-state) of $H$. By examining the shape of $V$, we
can guess a reasonable trial function $\Phi$, which approximates
$\Psi$ when $\Psi$ is large (i.e., near $r=1$). By taking the
Laplacian of $\Phi$, we can cast $\Phi$ as the groundstate of a
different Schroedinger equation.

Define
$$
U(r)-E_0\equiv \Phi^{-1}(\frac{1}{2}{\bf \nabla}^2 \Phi)\eqno(1.6)
$$
and
$$
H_0\equiv -\frac{1}{2}{\bf \nabla}^2 +U(r),\eqno(1.7)
$$
then
$$
H_0\Phi=E_0\Phi.\eqno(1.8)
$$
Introducing
$$
w \equiv H_0-H=U-V\eqno(1.9)
$$
and
$$
{\cal E}=E_0-E\eqno(1.10)
$$
the original Schroedinger equation (1.1) can be written as
$$
(H_0-E_0)\Psi=(w-{\cal E})\Psi.\eqno(1.11)
$$
In (1.6), only the difference $U-E_0$ is defined. The constant
$E_0$ may be chosen by requiring
$$
w(\infty)=0.\eqno(1.12)
$$

Following Refs.[11-14], the original Schroedinger equation (1.1)
will be solved through an iterative sequence
$$
(H_0-E_0)\Psi(m)=(w-{\cal E}_m)\Psi(m-1)\eqno(1.13)
$$
with $m=1,~2,~\cdots$ and when $m=0$
$$
\Psi(0)=\Phi.\eqno(1.14)
$$
The following simple discussion provides the motivation of this
approach. Multiplying (1.13) by $\Phi$ and (1.8) by $\Psi(m)$, we
find their difference given by the familiar Wronskian-type
expression
$$
-\frac{1}{2}{\bf \nabla}\cdot(\Phi {\bf \nabla}\Psi(m)-\Psi(m)
{\bf \nabla}\Phi)=(w-{\cal E}_m)\Phi\Psi(m-1).\eqno(1.15)
$$
Integrating (1.15) over all space, we have
$$
{\cal E}_m=\int w\Phi\Psi(m-1)d^Nq\bigg{/}\int
\Phi\Psi(m-1)d^Nq~.\eqno(1.16)
$$

Note that if $\Psi^0(m)$ is a solution of (1.13), so is
$$
\Psi(m)=\Psi^0(m)+c~\Phi,\eqno(1.17)
$$
where $c$ is a constant. Since $\Phi$ is the groundstate of $H_0$,
it is positive everywhere. By adjusting the constant $c$,
$\Psi(m)$ can also be made positive everywhere. Thus, if $w$ is
bounded, so is ${\cal E}_m$. This approach prevents the kind of
divergence encountered by the usual perturbative series, and
enables us to derive explicit convergent iterative solutions, as
we shall see.

In Section 2, we show that when $V$ is a radial potential, the
successive iterative solutions $\Psi(m)$ can be solved by simple
quadratures. For the Sombrero-shaped potential (1.4) the low-lying
wave function $\Psi$ is known reasonably well when it is in the
asymptotic region when $r$ is large, and also when its amplitude
is large (i.e., $r$ near $1$). In Section 3, a reasonable trial
function $\Phi$ is constructed whose fractional deviation from
$\Psi$ is large only for $r$ small (i.e., when the absolute
magnitude of $\Psi$ is also very small); the same should also
apply to $w(r)$, the difference between $H_0$ and $H$, given by
(1.9). As will be proved in Section 4, $w(r)$ is positive
everywhere and its radial derivative $w'(r)$ always negative,
making $w$ maximal at $r=0$. These conditions enable us to apply
the Hierarchy Theorem[12-14] which guarantees the convergence, as
will be discussed in Section 5. In Appendix B, we show that the
rate of convergence for large $g$ is similar to a power series in
$g^{-1}$.

\newpage

\section*{\bf 2. The Angular Momentum Operator and the Radial Equation}
\setcounter{section}{2} \setcounter{equation}{0}

Express the Cartesian coordinates $q_1,~q_2,~\cdots,~q_N$ of (1.2)
in terms of the radial variable $r$ and the $(N-1)$ angular
variables
\begin{eqnarray}\label{e3.1}
\theta_1,~\theta_2,\cdots,~\theta_{N-2}~~{\sf and}~~\theta_{N-1}
\end{eqnarray}
through
\begin{eqnarray}\label{e3.2}
q_1&=&r \cos \theta_1,~~~~q_2=r \sin \theta_1 \cos
\theta_2,\nonumber\\
q_3&=&r \sin \theta_1 \sin \theta_2 \cos
\theta_3, \cdots, \\
q_{N-1}&=& r \sin \theta_1 \sin \theta_2 \cdots \sin \theta_{N-2}
\cos \theta_{N-1}\nonumber
\end{eqnarray}
and
\begin{eqnarray*}
~~~q_N~=~r \sin\theta_1 \sin \theta_2 \cdots \sin \theta_{N-2}
\sin \theta_{N-1}
\end{eqnarray*}
with
\begin{eqnarray}\label{e3.3}
0 &\leq& \theta_i< \pi ~~{\sf
for}~~i=1,~2,\cdots,~N-2\nonumber\\
{\sf and}~~~~~~~~~~~~~~&&\\
0&\leq& \theta_{N-1}\leq 2 \pi.\nonumber
\end{eqnarray}
Correspondingly, the line elements are
\begin{eqnarray}\label{e3.4}
dr,~rd\theta_1,~r\sin \theta_1 d\theta_2,~r \sin \theta_1
 \sin \theta_2 d \theta_3,~~~~~~~~~~~~~~\nonumber\\
 r \sin\theta_1 \sin \theta_2 \sin \theta_3 d\theta_4,~\cdots,
 ~r \sin \theta_1 \sin\theta_2 \cdots \sin \theta_{N-2}d\theta_{N-1}
\end{eqnarray}
and the Laplacian operator is
\begin{eqnarray}\label{e3.5}
\nabla^2=\frac{1}{r^{2K}}~\frac{\partial}{\partial
r}(r^{2K}\frac{\partial}{\partial r}) -\frac{1}{r^2} {\cal
L}^2(N-1)
\end{eqnarray}
where
\begin{eqnarray*}
K=\frac{1}{2}(N-1),
\end{eqnarray*}
\begin{eqnarray}\label{e3.6}
{\cal L}^2(N-1)&=&-\frac{1}{\sin^{N-2}
\theta_1}\frac{\partial}{\partial \theta_1}(\sin^{N-2}
\theta_1\frac{\partial}{\partial \theta_1}) +\frac{1}{\sin^2 \theta_1} {\cal L}^2(N-2) \nonumber\\
{\cal L}^2(N-2)&=&-\frac{1}{\sin^{N-3}
\theta_2}\frac{\partial}{\partial \theta_2}(\sin^{N-3}
\theta_2\frac{\partial}{\partial \theta_2}) +\frac{1}{\sin^2 \theta_2} {\cal L}^2(N-3) \nonumber\\
\cdots&&\cdots\\
{\cal L}^2(2)&=&-\frac{1}{\sin
\theta_{N-2}}\frac{\partial}{\partial \theta_{N-2}}(\sin
\theta_{N-2}\frac{\partial}{\partial \theta_{N-2}})
+\frac{1}{\sin^2 \theta_{N-2}} {\cal L}^2(1) \nonumber
\end{eqnarray}
and
\begin{eqnarray*}
 {\cal L}^2(1)=-\frac{\partial^2}{\partial \theta^2_{N-1}}.
\end{eqnarray*}
The square of the angular momentum operator on an $n$-sphere is
${\cal L}^2(n)$. From (\ref{e3.6}), one sees readily that the
commutator between any two ${\cal L}^2(n)$ and ${\cal L}^2(m)$ is
zero; i.e.,
\begin{eqnarray}\label{e3.7}
 [{\cal L}^2(n),~{\cal L}^2(m)]=0.
\end{eqnarray}

As we shall show in Appendix A, the eigenvalues of each ${\cal
L}^2(n)$ are
\begin{eqnarray}\label{e3.8}
l(l+n-1)
\end{eqnarray}
with $l=0,~1,~2,\cdots$. Thus, for a radially symmetric potential,
the wave function can be written as
\begin{eqnarray}\label{e3.9}
\Psi(r,~\theta_1,~\theta_2,\cdots,~\theta_{N-1})={\cal R}(r)
\Theta (\theta_1,~\theta_2,\cdots,~\theta_{N-1})
\end{eqnarray}
with
\begin{eqnarray}\label{e3.10}
{\cal L}^2(N-1) \Theta &=&l_1(l_1+N-2) \Theta, \nonumber\\
{\cal L}^2(N-2) \Theta&=&l_2(l_2+N-3) \Theta, \nonumber\\
&&\cdots\\
{\cal L}^2(2) \Theta&=&l_{N-2}(l_{N-2}+1) \Theta\nonumber
\end{eqnarray}
and
\begin{eqnarray*}
{\cal L}^2(1) \Theta=l_{N-1}^2 \Theta.
\end{eqnarray*}
Correspondingly, by using (2.5) we find the radial part of the
Schroedinger equation (1.1) for the lowest eigenstate of angular
momentum $l$ to be
\begin{eqnarray}\label{e3.11}
[-\frac{1}{2}\nabla_r^2 + \frac{1}{2r^2}l(l+N-2)+V(r)-E~]~{\cal
R}(r)=0
\end{eqnarray}
with
\begin{eqnarray}\label{e3.12}
 \nabla_r^2 = \frac{1}{r^{2K}}\frac{d}{dr}(r^{2K}\frac{d}{dr})
\end{eqnarray}
and $l=l_1$ given by the first equation of (\ref{e3.10}); i.e.,
\begin{eqnarray}\label{e3.13}
{\cal L}^2(N-1) \Theta = l(l+N-2) \Theta.
\end{eqnarray}

Define
$$
{\cal R}(r)=r^{-K}\psi(r).\eqno(2.14)
$$
Eq.(\ref{e3.11}) becomes
$$
[-\frac{1}{2}\frac{d^2}{dr^2}+\frac{1}{2r^2}k(k-1))+
V(r)-E]\psi(r)=0. \eqno(2.15)
$$
where
$$
k=l+K=l+\frac{1}{2}(N-1)\eqno(2.16)
$$
and therefore $k(k-1)=K(K-1)+l(l+N-2)$, on account of
$K=\frac{1}{2}(N-1)$. As $r\rightarrow 0$, ${\cal R} \propto r^l$
and therefore $\psi \propto r^{l+K}=r^k$.

The radial trial function $\phi(r)$ shall be constructed in the
next section. Similar to (2.15) it satisfies
$$
(H_0-E_0)\phi(r)=0\eqno(2.17)
$$
with
$$
H_0=-\frac{1}{2}\frac{d^2}{dr^2}+\frac{1}{2r^2}k(k-1)+U(r).\eqno(2.18)
$$
The iterative equation (1.11) becomes
$$
(H_0-E_0)\psi_m(r)=(w(r)-{\cal E}_m)\psi_{m-1}(r)\eqno(2.19)
$$
where, as before,
$$
w(r)=U(r)-V(r).\eqno(2.20)
$$
Define
$$
f_m(r)\equiv \psi_m(r)/\phi(r).\eqno(2.21)
$$
As in (1.15)-(1.16), we have
$$
-\frac{1}{2}\frac{d}{dr}(\phi^2\frac{df_m}{dr})=(w-{\cal
E}_m)\phi^2f_{m-1}\eqno(2.22)
$$
and
$$
{\cal E}_m=\int\limits_0^{\infty}
w\phi^2f_{m-1}dr\bigg{/}\int\limits_0^{\infty}
\phi^2f_{m-1}dr.\eqno(2.23)
$$
Eq. (2.22) can be readily integrated
$$
f_m(r)=f_m(\infty)-2\int\limits_r^{\infty} \frac{dy}{\phi^2(y)}
\int\limits_{y}^{\infty} (w(x)-{\cal
E}_m)\phi^2(x)f_{m-1}(x)dx\eqno(2.24)
$$
or, equivalently
$$
f_m(r)=f_m(0)-2\int\limits_0^{r} \frac{dy}{\phi^2(y)}
\int\limits_0^{y} (w(x)-{\cal
E}_m)\phi^2(x)f_{m-1}(x)dx.\eqno(2.25)
$$
Thus, with
$$
f_0(r)=1,
$$
each $f_m(r)$ can be derived by quadratures. At each iteration,
there is an arbitrary constant $f_m(\infty)$ or $f_m(0)$ (similar
to the constant $c$ in (1.17)). As we shall see, different choices
of this constant can lead to different convergent iterative
solutions.\\

\noindent Remarks: We note that in (2.11), the centrifugal
potential $\frac{1}{2r^2}l(l+N-2)$ becomes the familiar $l^2/2r^2$
in two-dimension and $l(l+1)/2r^2$ in three-dimension.
Correspondingly, the term $k(k-1)/2r^2$ in (2.15) becomes
$(l^2-\frac{1}{4})/2r^2$ in two-dimension, but remains
$l(l+1)/2r^2$ in three-dimension.

\newpage

\section*{\bf 3. The Trial Function }
\setcounter{section}{3} \setcounter{equation}{0}

To construct a reasonable trial function $\phi(r)$, we start with
the following two functions $\phi_+(r)$ and $\phi_-(r)$.
\begin{eqnarray}\label{e1.7}
\phi_+(r) = \frac{2r^k}{r+1}(\frac{1+a}{r+a})^k~e^{-gS_0(r)}
\end{eqnarray}
and
\begin{eqnarray}\label{e1.8}
\phi_-(r) =
\frac{2r^k}{r+1}(\frac{1+a}{r+a})^k~e^{-\frac{4}{3}g+gS_0(r)}
\end{eqnarray}
where
\begin{eqnarray}\label{e1.9}
S_0(r) = \frac{1}{3}(r-1)^2(r+2)
\end{eqnarray}
and, as before, $k$ is related to the angular momentum $l$ and the
dimensionality $N$ by
\begin{eqnarray}\label{e1.10}
k = l+ \frac{1}{2}(N-1).
\end{eqnarray}
The parameter $a$ is positive but free within a range, which will
be specified below. The trial function $\phi(r)$ is given by
\begin{eqnarray}\label{e1.11}
\phi (r) = \left\{
\begin{array}{ll}
g_+\phi_+(r)+g_-\phi_-(r)
&~~~~~~{\sf for}~~0 \leq r<1\\
(g_++g_-e^{-\frac{4}{3}g})\phi_+(r)&~~~~~~{\sf for}~~r>1,
\end{array}
\right.
\end{eqnarray}
where $g_+$ and $g_-$ are constants given by
\begin{eqnarray}\label{e1.12}
g_\pm= g\pm (\frac{k}{a}+1).
\end{eqnarray}
Thus, by construction $r^{-k}\phi$ is regular at $r=0$, with its
derivative
\begin{eqnarray}\label{e1.13}
(r^{-k}\phi)'\equiv \frac{d}{dr}(r^{-k}\phi)=0~~~{\sf at}~~r=0.
\end{eqnarray}
Throughout, ' denotes $d/dr$. By construction, $\phi(r)$ and
$\phi'(r)$ are continuous everywhere, with $r$ varying from $0$ to
$\infty$.

By differentiation, $\phi_+(r)$ and $\phi(r)$ satisfy respectively
\begin{eqnarray}\label{e1.14}
(-\frac{1}{2}\frac{d^2}{dr^2}+\frac{1}{2r^2}k(k-1) + V+h )\phi_+
=E_0\phi_+
\end{eqnarray}
and, as in (2.17)-(2.20),
\begin{eqnarray}\label{e1.15}
( -\frac{1}{2}\frac{d^2}{dr^2}+\frac{1}{2r^2}k(k-1)+ V+w )\phi
=E_0\phi,
\end{eqnarray}
where
\begin{eqnarray}\label{e1.16}
E_0=g(1+ka),
\end{eqnarray}
\begin{eqnarray}\label{e1.17}
h(r)=\frac{kag_-}{r(r+a)}+\frac{ka^2g}{r+a}+\frac{1}{(r+1)^2}
+\frac{k(k+1)}{2(r+a)^2}+\frac{ka}{(r+a)(r+1)}
\end{eqnarray}
and
\begin{eqnarray}\label{e1.18}
w(r)=h(r)+\hat{g}(r)
\end{eqnarray}
with
\begin{eqnarray}\label{e1.19}
\hat{g} (r) = \left\{\begin{array}{ll}
\frac{2gg_-[1-\frac{ka(1-r^2)}{r(r+a)}]}{g_+e^{\Lambda} + g_-},
&~~~~~~{\sf for}~~0 \leq r<1\\
0&~~~~~~{\sf for}~~r>1,
\end{array}
\right.
\end{eqnarray}
where
\begin{eqnarray}\label{e1.20}
\Lambda=2g(r-\frac{r^3}{3})
\end{eqnarray}
and by construction $[g_-/(g_+e^\Lambda+g_-)]=g_-\phi_-/\phi$.

\noindent \underline{Remarks}

\noindent (i) The exponent $\mp gS_0(r)$ in (\ref{e1.7}) and
(\ref{e1.8}) satisfies
\begin{eqnarray}\label{e1.21}
\frac{1}{2}[\pm gS'_0(r)]^2=\frac{1}{2}g^2(r^2-1)^2=V(r).
\end{eqnarray}
(ii) For $l=0$ and $N=1$, we have $k=0$. Correspondingly,
$\phi_+(r)$ of (\ref{e1.7}) becomes
\begin{eqnarray}\label{e1.22}
\phi_0(r) = \frac{2}{r+1}~e^{-gS_0(r)}
\end{eqnarray}
which satisfies
\begin{eqnarray}\label{e1.23}
( -\frac{1}{2} \frac{d^2}{dr^2} + V+h_0 )\phi_0 =g\phi_0
\end{eqnarray}
with
\begin{eqnarray}\label{e1.24}
h_0(r)=\frac{1}{(r+1)^2}.
\end{eqnarray}
These equations reduce to the same expressions used before for the
corresponding one-dimensional trial function[12, 13].

(iii) From (3.11) and (3.13), we see that at $r=0$ both $h$ and
$\hat{g}$ contain an $r^{-1}$ pole term. This is because neither
the derivative of $r^{-k}\phi_+(r)$ nor that of $r^{-k}\phi_-(r)$
is zero at $r=0$. The trial function $\phi(r)$ does satisfy
$(r^{-k}\phi)'=0$ at $r=0$, in accordance with (3.5) and (3.7).
Thus, the potential function $w(r)=h(r)+\hat{g}(r)$ given by
(3.12) is regular at $r=0$.

Throughout the paper, we assume
\begin{eqnarray}\label{e1.26}
g>\frac{k}{a}+1,
\end{eqnarray}
so that $g_->0$. Correspondingly, as will be shown, $w(r)$ is
positive and has a discontinuity at $r=1$.

\newpage

\section*{\bf 4. Properties of $w(r)$}
\setcounter{section}{4} \setcounter{equation}{0}

In this section, we shall establish
\begin{eqnarray}\label{e2.1}
w(r)&>&0 \nonumber\\
{\sf and}~~~~~~~~~~~~~~~~~~~~~~~~~~~~~~~~~~~&&\\
w'(r)&<&0~~~{\sf for}~~~r\geq
0.~~~~~~~~~~~~~~~~~~~~~~~~~~~~~~~~\nonumber
\end{eqnarray}
It is convenient to write (\ref{e1.17}) as
\begin{eqnarray}\label{e2.2}
h(r) = \sum_{i=1}^{5} h_i(r)
\end{eqnarray}
with
\begin{eqnarray}\label{e2.3}
h_1(r)=\frac{kag_-}{r(r+a)},~~~h_2(r)=\frac{ka^2g}{r+a},~~~h_3(r)=\frac{1}{(r+1)^2},\nonumber\\
h_4(r)=\frac{k(k+1)}{2(r+a)^2}~~~{\sf
and}~~~h_5(r)=\frac{ka}{(r+a)(r+1)}.
\end{eqnarray}
Likewise, we decompose (\ref{e1.19}) for $r<1$ as
\begin{eqnarray}\label{e2.4}
\hat{g}(r) = \sum_{i=6}^{8} \hat{g}_i(r)
\end{eqnarray}
with
\begin{eqnarray}\label{e2.5}
\hat{g}_6(r)&=&\frac{2gg_-}{g_+e^\Lambda
+g_-},~~~\hat{g}_7(r)=\frac{2gg_-kar}{(g_+e^\Lambda
+g_-)(r+a)}~~~~~~~~~~~~~~\nonumber\\
{\sf
and}~~~~~~~~~~~~~~~~&&~~~\hat{g}_8(r)=-\frac{2kgg_-a}{(g_+e^\Lambda
+g_-)r(r+a)}.
\end{eqnarray}
In order to show that $w=h+\hat{g}$ satisfies (\ref{e2.1}), we
note that for $r>1$, $\hat{g}=0$ in accordance with (\ref{e1.9}),
and therefore
\begin{eqnarray}
w(r)=h(r).\nonumber
\end{eqnarray}
Since each of the $h_i(r)$ in (\ref{e2.3}) satisfies
\begin{eqnarray}
h_i(r)>0~~~{\sf and}~~~h'_i(r)<0,\nonumber
\end{eqnarray}
therefore
\begin{eqnarray}\label{e2.6}
w(r)>0~~~{\sf and}~~~w'(r)<0~~~{\sf for}~~r>1.
\end{eqnarray}
For $r<1$, it is convenient to combine first some of the $h_i$
with $\hat{g}_j$ given by (\ref{e2.3}) and (\ref{e2.5}).

Defining
\begin{eqnarray}\label{e2.7}
w_I \equiv h_1+\hat{g}_8,
\end{eqnarray}
we find
\begin{eqnarray}\label{e2.8}
w_I =\frac{ka}{r(r+a)}(g_--\frac{2gg_-}{g_+e^\Lambda +g_-})
=\frac{kag_+g_-(e^\Lambda-1)}{r(r+a)(g_+e^\Lambda+g_-)}~.
\end{eqnarray}
Likewise, write
\begin{eqnarray}\label{e2.9}
h_2+\hat{g}_7 &=& gka\bigg[\frac{a}{r+a}+\frac{2g_-r}{(g_+e^\Lambda +g_-)(r+a)}\bigg]\nonumber\\
&=&gka \bigg\{\frac{a}{r+a}+\frac{2g_-[(r+a)-a]}{(g_+e^\Lambda+g_-)(r+a)}\bigg\}\nonumber\\
&=&gka
\bigg\{\frac{a}{r+a}[1-\frac{2g_-}{g_+e^\Lambda+g_-}]+\frac{2g_-}{g_+e^\Lambda+g_-}\bigg\}.
\end{eqnarray}
Separate the two terms inside the curly brackets and write
\begin{eqnarray}\label{e2.10}
h_2+\hat{g}_7=w_{II}+w_0
\end{eqnarray}
with
\begin{eqnarray}\label{e2.11}
w_{II}
=\frac{gka^2}{r+a}\bigg[\frac{g_+e^\Lambda-g_-}{g_+e^\Lambda+g_-}\bigg]
\end{eqnarray}
in which the factor inside the square brackets is identical to the
corresponding one in (\ref{e2.9}), and
\begin{eqnarray}\label{e2.12}
w_0 =\frac{2gg_-ka} {g_+e^\Lambda+g_-}.
\end{eqnarray}
Next, combine the above $w_0$ with $\hat{g}_6$ of (\ref{e2.5})
into a single term as follows:
\begin{eqnarray}\label{e2.13}
w_{VI} \equiv w_0 +\hat{g}_6
=\frac{2gg_-(ka+1)}{g_+e^\Lambda+g_-}.
\end{eqnarray}

We now express $w=h+\hat{g}$ , for $r<1$, in terms of a new sum of
six terms:
\begin{eqnarray}\label{e2.14}
w = \sum_{m=I}^{VI} w_m
\end{eqnarray}
with $w_I$, $w_{II}$ and $w_{VI}$ given by (\ref{e2.9}),
(\ref{e2.11}) and (\ref{e2.13}) respectively; the rest
$w_{III}=h_3$, $w_{IV}=h_4$ and $w_V=h_5$. In explicit forms
\begin{eqnarray}\label{e2.15}
w_I &=&
\frac{Zka}{r(r+a)},~~~~~~~~~~~~~~~~~w_{II}=\frac{Ygka^2}{r+a},\nonumber\\
w_{III}&=&\frac{1}{(r+1)^2},~~~~~~~~~~~~~~~~~w_{IV}=\frac{k(k+1)}{2(r+a)^2},\\
w_V&=&\frac{ka}{(r+a)(r+1)},~~{\sf
and}~~~w_{VI}=X2g(ka+1)\nonumber
\end{eqnarray}
where
\begin{eqnarray}\label{e2.16}
X =\frac{g_-} {g_+e^\Lambda+g_-},
\end{eqnarray}
\begin{eqnarray}\label{e2.17}
Y =\frac{g_+e^\Lambda-g_-} {g_+e^\Lambda+g_-}.
\end{eqnarray}
and
\begin{eqnarray}\label{e2.18}
Z =\frac{g_+g_-(e^\Lambda-1)} {g_+e^\Lambda+g_-}.
\end{eqnarray}
For the $w_m$ with $m={\tiny III,~IV}$ and $V$, it is clear that
\begin{eqnarray}\label{e2.19}
w_m>0~~~{\sf and}~~~w'_m<0.
\end{eqnarray}
For $m=VI$, since in accordance with (\ref{e1.20}), for $r<1$,
$\Lambda>0$ and
\begin{eqnarray}\label{e2.20}
\Lambda'=2g(1-r^2)>0,
\end{eqnarray}
$w_{VI}$ also satisfies (\ref{e2.19}).

Introduce
\begin{eqnarray}\label{e2.21}
\xi =\frac{1}{2}\Lambda=g(r-\frac{r^3}{3})
\end{eqnarray}
and
\begin{eqnarray}\label{e2.22}
e^{2b}=\frac{g_+}{g_-}.
\end{eqnarray}
Recall
\begin{eqnarray}\label{e2.23}
g_+ + g_-=2g,
\end{eqnarray}
on account of (\ref{e1.12}). we can express $g_{\pm}$ in terms of
$g$ and $b$:
\begin{eqnarray}\label{e2.24}
g_+ =\frac{ge^b}{\cosh b}~~~{\sf and}~~~g_-=\frac{ge^{-b}}{\cosh
b}.
\end{eqnarray}
Thus, we can rewrite (\ref{e2.18}) as
\begin{eqnarray}\label{e2.25}
Z=\frac{g}{\cosh b}~\frac{\sinh \xi}{\cosh(\xi+b)},
\end{eqnarray}
and correspondingly
\begin{eqnarray}\label{e2.26}
\frac{Z}{r}= \frac{g}{\cosh
b}\bigg(\frac{\xi}{r}\bigg)\bigg(\frac{\tanh
\xi}{\xi}\bigg)\bigg(\frac{\cosh \xi}{\cosh(\xi+b)}\bigg).
\end{eqnarray}
We observe that
\begin{eqnarray}\label{e2.27}
\frac{d}{dr}\bigg(\frac{\xi}{r}\bigg)=-g\frac{2r}{3}<0
\end{eqnarray}
and for $r<1$,
\begin{eqnarray}\label{e2.28}
\frac{d\xi}{dr}=g(1-r^2)>0.
\end{eqnarray}
Since for $\xi$ positive
\begin{eqnarray}\label{e2.29}
\frac{d}{d\xi}\bigg(\frac{\tanh \xi}{\xi}\bigg)=
\frac{1}{2\xi^2\cosh^2 \xi}(2\xi-\sinh 2\xi)<0
\end{eqnarray}
and
\begin{eqnarray}\label{e2.30}
\frac{d}{d\xi}\ln \bigg(\frac{\cosh \xi}{\cosh (\xi+b)}\bigg)=
\tanh \xi-\tanh (\xi+b)<0,
\end{eqnarray}
we have for $r<1$
\begin{eqnarray}\label{e2.31}
\frac{d}{dr}\bigg(\frac{Z}{r}\bigg)<0.
\end{eqnarray}
Thus from (\ref{e2.15}),
\begin{eqnarray}
w_I>0\nonumber
\end{eqnarray}
and
\begin{eqnarray}\label{e2.32}
w'_I<0~~~{\sf for}~~~r<1.
\end{eqnarray}

Lastly, we examine $w_{II}$. From (\ref{e2.17}) we see that
\begin{eqnarray}\label{e2.33}
Y>0,
\end{eqnarray}
and therefore, in accordance with (\ref{e2.15})
\begin{eqnarray}\label{e2.34}
w_{II}=\frac{Ygka^2}{r+a}>0.
\end{eqnarray}
By using (\ref{e2.17}) and (\ref{e2.20}), we also have
\begin{eqnarray}\label{e2.35}
(\ln Y)'
&=&\bigg(\frac{1}{g_+e^\Lambda-g_-}-\frac{1}{g_+e^\Lambda+g_-}\bigg)
g_+e^\Lambda2g(1-r^2)\nonumber\\
&=&\bigg(\frac{2}{e^{2(\Lambda+2b)}-1}\bigg)e^{\Lambda+2b}2g(1-r^2)\nonumber\\
&=&\frac{2g(1-r^2)}{\sinh (\Lambda+2b)}>0
\end{eqnarray}
for $r<1$. Therefore
\begin{eqnarray}\label{e2.36}
(\ln w_{II})'&=&(\ln Y)'-\frac{1}{r+a} \nonumber\\
&=&\frac{2g(1-r^2)}{\sinh (\Lambda+2b)}-\frac{1}{r+a}\nonumber\\
&=&\frac{1}{(r+a)\sinh (\Lambda+2b)}[T-S]
\end{eqnarray}
with
\begin{eqnarray}\label{e2.37}
S\equiv \sinh (\Lambda+2b)
\end{eqnarray}
and
\begin{eqnarray}\label{e2.38}
T\equiv 2g(1-r^2)(r+a).
\end{eqnarray}
We seek to show that $T-S<0$ for $0<r<1$. On account of
(\ref{e2.20}),
\begin{eqnarray}
(T-S)'&=&-4gr(r+a)+2g(1-r^2)-\Lambda' \cosh(\Lambda+2b)\nonumber\\
&=&-4gr(r+a)+2g(1-r^2)[1- \cosh(\Lambda+2b)].\nonumber
\end{eqnarray}
At any $r<1$, $(T-S)'<0$ and therefore
\begin{eqnarray}\label{e2.39}
T(r)-S(r)<T(0)-S(0).
\end{eqnarray}
When $r=0$, $\Lambda(0)=0$,
\begin{eqnarray}\label{e2.40}
S(0)=\sinh
2b&=&\frac{1}{2}\bigg(\frac{g_+}{g_-}-\frac{g_-}{g_+}\bigg)
=\frac{g_+^2-g_-^2}{2g_+g_-}\nonumber\\
&=&\frac{1}{2}\bigg[\frac{(g+\frac{k}{a}+1)^2-(g-\frac{k}{a}-1)^2}
{g^2-(\frac{k}{a}+1)^2}\bigg]\nonumber\\
&=&\frac{2g(\frac{k}{a}+1)}{g^2-(\frac{k}{a}+1)^2}
\end{eqnarray}
and
\begin{eqnarray}\label{e2.41}
T(0)=2ga.
\end{eqnarray}
Setting the parameter $a$ to be within an upper bound defined by
\begin{eqnarray}\label{e2.42}
a<\frac{\frac{k}{a}+1}{g^2-(\frac{k}{a}+1)^2},
\end{eqnarray}
we have $T(0)-S(0)<0$, and therefore
\begin{eqnarray}\label{e2.43}
w_{II}'(0)<0~~~{\sf for}~~~r<1.
\end{eqnarray}
The inequality (\ref{e2.42}) can also be written as
\begin{eqnarray}\label{e2.48}
g^2<(\frac{k}{a}+1) (\frac{k}{a}+\frac{1}{a}+1).
\end{eqnarray}
Combining with the inequality (\ref{e1.26}), we require
\begin{eqnarray}\label{e2.49}
1<\frac{g}{\frac{k}{a}+1}<\sqrt{1+\frac{1}{k+a}}.
\end{eqnarray}
This is the sufficient condition for $w_{II}'<0$, and therefore
also
\begin{eqnarray}\label{e2.50}
w'<0~~~{\sf for}~~~r<1.
\end{eqnarray}
At $r=1$, in accordance with (\ref{e2.4})-(\ref{e2.5})
\begin{eqnarray}\label{e2.51}
\hat{g}_7(1)+\hat{g}_8(1)=0
\end{eqnarray}
and therefore
\begin{eqnarray}\label{e2.52}
\hat{g}(1)=\hat{g}_6(1)=\frac{2gg_-}{g_+e^{\frac{4}{3}g}+g_-}>0.
\end{eqnarray}
Thus, $w(r)$ has a discontinuity at $r=1$, with
\begin{eqnarray}\label{e2.53}
w(1-)-w(1+)=\hat{g}(1)>0
\end{eqnarray}
Together with (\ref{e2.6}) and(\ref{e2.50}), this proves that for
the parameter $a$ within the range (\ref{e2.49}),
\begin{eqnarray}
w(r)>0~~~{\sf and}~~~w'(r)<0\nonumber
\end{eqnarray}
over the entire range of $r$.\\

\noindent \underline{Remarks}

\noindent (i) The first inequality in(\ref{e2.49})
\begin{eqnarray}
\frac{k}{a}+1<g\nonumber
\end{eqnarray}
prevents the limit $a\rightarrow 0$; otherwise $h(r)$ would have a
double pole $r^{-2}$, in accordance with (4.2)-(4.3).

\noindent (ii) Choose the parameter $a$ within the range
(\ref{e2.49}). As we will discuss in the next section, because of
(\ref{e2.1}), the application of the Hierarchy Theorem leads to a
convergent iterative solution in terms of quadratures for the
lowest eigenstate Schroedinger wave function (1.1)-(1.5) in any
dimension $N$ and with arbitrary angular momentum $l$.

\newpage

\section*{\bf 5. The Convergent Iterative Solution}
\setcounter{section}{5} \setcounter{equation}{0}

\noindent {\bf 5.1 The Hierarchy Theorem}\\

We begin with the iterative equations (2.22)-(2.25). Recalling
that $f_m(r)$ and ${\cal E}_m$ are the $m^{th}$ order solutions
for $f=\psi/\phi$ and ${\cal E}=E_0-E$, we derive the
corresponding $m^{th}$ order solutions for $\psi$ and $E$ to be
$$
\psi_m=\phi f_m~~~~~~~{\sf and}~~~~~~E_m=E_0-{\cal E}_m\eqno(5.1)
$$
Throughout this section, we assume the parameter $a$ to be
restricted by the two inequalities given by (4.45), Therefore, for
large $g$, $a$ is small. We distinguish two different conditions:
$$
(A)~~f_m(\infty)=1~~~{\sf for~all}~~m\eqno(5.2)
$$
or
$$
(B)~~f_m(0)=1~~~{\sf for~all}~~m.\eqno(5.3)
$$
Correspondingly in $(A)$, (2.24) becomes
$$
f_m(r)=1-2\int\limits_r^{\infty} \frac{dy}{\phi^2(y)}
\int\limits_{y}^{\infty} (w(x)-{\cal
E}_m)\phi^2(x)f_{m-1}(x)dx\eqno(5.4)
$$
and in $(B)$, (2.25) becomes
$$
f_m(r)=1-2\int\limits_0^{r} \frac{dy}{\phi^2(y)} \int\limits_0^{y}
(w(x)-{\cal E}_m)\phi^2(x)f_{m-1}(x)dx.\eqno(5.5)
$$
In both cases, we assume that the two inequalities of (4.45) hold.
Thus $w(r)$ is positive and $w'(r)$ negative at all $r$, in
accordance with (4.1). In case $(A)$, apart from these two
inequalities of (4.45) there is no other restriction on the
magnitude of $w(r)$. In case $(B)$, we assume $w(r)$ to be not too
large so that for all $m$
$$
f_m(r)>0~~~~~{\sf at~~all}~~r.\eqno(5.6)
$$
As we shall see, this is equivalent to requiring for all $m$
$$
f_m(\infty)>0.\eqno(5.7)
$$

\noindent \underline{Hierarchy Theorem}[12-14]\\

$(A)$ With the boundary condition $f_m(\infty)=1$, we have for all
$m$
$$
{\cal E}_{m+1} > {\cal E}_m\eqno(5.8)
$$
and
$$
\frac{d}{dr}\left( \frac{f_{m+1}(r)}{f_m(r)}\right)<0 ~~~~~{\sf
at~~any}~~~r>0.\eqno(5.9)
$$
Thus, the sequences $\{{\cal E}_m\}$ and $\{f_m(r)\}$ are all
monotonic, with
$$
{\cal E}_1<{\cal E}_2<{\cal E}_3<\cdots\eqno(5.10)
$$
and
$$
1<f_1(r)<f_2(r)<f_3(r)<\cdots\eqno(5.11)
$$
at all finite $r$.

$(B)$ With the boundary condition $f_m(0)=1$, we have for all odd
$m=2n+1$ an ascending sequence
$$
{\cal E}_1<{\cal E}_3<{\cal E}_5<\cdots,\eqno(5.12)
$$
but for all even $m=2n$, a descending sequence
$$
{\cal E}_2>{\cal E}_4>{\cal E}_6>\cdots.\eqno(5.13)
$$
In addition, between any even $m=2n$ and any odd $m=2l+1$
$$
{\cal E}_{2n}>{\cal E}_{2l+1}.\eqno(5.14)
$$
Likewise, at any $r$, for any even $m=2n$
$$
\frac{d}{dr}\left(
\frac{f_{2n+1}(r)}{f_{2n}(r)}\right)<0,\eqno(5.15)
$$
whereas for any odd $m=2l+1$
$$
\frac{d}{dr}\left(
\frac{f_{2l+2}(r)}{f_{2l+1}(r)}\right)>0.\eqno(5.16)
$$
Furthermore,
$$
\lim\limits_{m\rightarrow \infty} E_m=E\eqno(5.17)
$$
and
$$
\lim\limits_{m\rightarrow \infty}
f_m(r)=f(r)=\psi(r)/\phi(r).\eqno(5.18)
$$
Thus, the boundary condition $f_m(\infty)=1$ yields a sequence, in
accordance with (5.10),
$$
E_1>E_2>E_3>\cdots>E,\eqno(5.19)
$$
with each member $E_m$ an {\it upper} bound of $E$, similar to the
usual variational method.

On the other hand, with the boundary condition $f_m(0) =1$, while
the sequence of its odd members $m=2l+1$ yields a similar one,
like (5.19), with
$$
E_1>E_3>E_5> \cdots >E,\eqno(5.20)
$$
its even members $m=2n$ satisfy
$$
E_2<E_4<E_6< \cdots <E.\eqno(5.21)
$$
It is unusual to have an iterative sequence of {\it lower} bounds
of the eigenvalue $E$. Together,  these sequences may be quite
efficient to pinpoint the limiting $E$.

In Appendix B, by examining a simple prototype example in this
class of problems, we shall show that for the Sombrero-shaped
potential when $g$ is large, the rate of convergence is similar to
a power series in $g^{-1}$.\\

\noindent {\bf 5.2 Numerical Results}\\

From (4.45), we see that for a given pair $(k,~g)$, in order to
apply the Hierarchy Theorem, the parameter $a$ should be within a
range
$$
a_{min}<a<a_{max}\eqno(5.22)
$$
with the limits $a_{min}$ and $a_{max}$ determined by
$$
g=1+\frac{k}{a_{min}}\eqno(5.23)
$$
and
$$
g^2=(1+\frac{k}{a_{max}})(1+\frac{k+1}{a_{max}}).\eqno(5.24)
$$
Alternatively, for a given pair $(k,~a)$, $g$ should be within
$$
g^2_{min}<g^2<g^2_{max}\eqno(5.25)
$$
with
$$
g_{min}=1+\frac{k}{a}\eqno(5.26)
$$
and
$$
g^2_{max}=(1+\frac{k}{a})(1+\frac{k+1}{a}).\eqno(5.27)
$$
Examples of these limiting values are given in Table 1.

Figure 1 gives examples of the iterative solutions of $\psi_n$ and
$E_n$ with different $n$ for the parameters $g=3$, $k=2$, and
$a=1.2$. One sees that for both boundary conditions $f_n(0)=1$ and
$f_n(\infty)=1$, the convergence sets in rapidly in accordance
with the Hierarchy Theorem. In the case of the boundary condition
$f(0)=1$, there is a limit to the range of these parameters, in
order that $f_n(\infty)>0$, in accordance with (5.7). For example
for $g=3$, the boundary condition $f(0)=1$ can be applied only for
$k\leq 2.5$, while the boundary condition $f(\infty)=1$ can be
applied to any values of $k$.

In Figures 2 and 3 we give the final radial wave function ${\cal
R}(r)$ and energy $E$ that satisfy (2.11) for
$$
g=3,~l=0~~{\sf and}~~N=3,~4,~5,~6,
$$
$$
{\sf and~~also~~for}~~~~~~~~~~~~~~~~~~~~~~~~~~~~~~~~~~~~~
~~~~~~~~~~~~~~~~~~~~~~~~~~~~~~~~~~~~~~~\eqno(5.28)
$$
$$
g=3,~N=3~~{\sf and}~~l=0,~1,~2,~3.
$$
Recalling (2.14) and (2.16), we have the curves in Fig.2 for $l=0$
and $N=3,~4,~5,~{\sf and}~6$ corresponding to
$k=K=\frac{N-1}{2}=1,~1.5,~2~{\sf and}~2.5$; ${\cal
R}(r)=\frac{1}{r^k}\psi(r)$. In Fig.3 the curves for $N=3$ and
$l=0,~1,~2,~{\sf and}~3$ correspond to $K=1$ and
$k=K+l=1,~2,~3,~{\sf and}~4$; ${\cal R}(r) =\frac{1}{r^K}\psi(r)
=\frac{1}{r}\psi(r)$.\\

\newpage

\section*{\bf References}

\noindent
[1] A. M. Polyakov, Nucl.Phys. B121 (1977), 429 \\
~[2] G. 't Hooft, in: A. Zichichi, Erice(Eds.), The why's of
subnuclear physics

\noindent ~~~~~~~~~~        Plenum, New York, 1977\\
~[3] E. Brezin, G. Parisi and J. Zinn-Justin, Phys.Rev. D16 (1977), 408\\
~[4] J. Zinn-Justin, J.Math.Phys. 22 (1981), 511 \\
~[5] J. Zinn-Justin, Nucl.Phys. B192 (1981), 125 \\
~[6] J. Zinn-Justin, in: J.-D. Zuber, R. Stora (Eds.), Recent
advances in field

\noindent ~~~~~~~~~theory and statistical mechanics,
Les Houches, session XXXIX, 1982 \\
~[7] J. Zinn-Justin, Private Communication \\
~[8] Sidney Coleman, Aspects of Symmetry, Press Syndicate of

\noindent ~~~~~~~~~the University of Cambridge, 1987  \\
~[9] E. Shuryak, Nucl.Phys. B302 (1988), 621

\noindent [10] S. V. Faleev and P. G. Silvestrov, Phys. Lett. A197
(1995), 372

\noindent [11] R. Friedberg, T. D. Lee, W. Q. Zhao and A. Cimenser

\noindent ~~~~~~~~~Ann.Phys. 294 (2001), 67\\
\noindent [12] R. Friedberg and T. D. Lee, Ann.Phys. 308 (2003),
263\\
\noindent [13] R. Friedberg and T. D. Lee, Ann. Phys. 316(2005)44\\
\noindent [14] T D. Lee, Journal of Statistical Physics (in press)

\newpage

\section*{\bf Appendix A}
\setcounter{section}{6} \setcounter{equation}{0}

Denote the solution $\Theta$ of (2.10) as
$$
\Theta = \Theta^{(N-1)}_{l_1 l_2 \cdots l_{N-1}}
(\theta_1,~\theta_2,\cdots,~\theta_{N-1}). \eqno(A.1)
$$
In this Appendix, we shall derive its explicit form inductively.

Write
$$
\Theta^{(N-1)}_{l_1 l_2 \cdots l_{N-1}}
(\theta_1,~\theta_2,\cdots,~\theta_{N-1})=Z^{(N)}_{l_1l_2}
\Theta^{(N-2)}_{l_2 l_3 \cdots l_{N-1}}
(\theta_2,~\theta_3,\cdots,~\theta_{N-1}) \eqno(A.2)
$$
with $Z^{(N)}_{l_1l_2}$ depending only on $\theta_1$. By using
(2.6) and (2.10), we see that $Z^{(N)}_{l_1l_2}$ satisfies
$$
\bigg[\frac{1}{\sin^{N-2} \theta_1}\frac{\partial}{\partial
\theta_1}(\sin^{N-2} \theta_1\frac{\partial}{\partial \theta_1})
+\frac{l_2(l_2+N-3)}{\sin^2 \theta_1} -l_1(l_1+N-2)\bigg]
Z^{(N)}_{l_1l_2}=0. \eqno(A.3)
$$
For $N>2$, it is convenient to denote
$$
z=\cos \theta_1,~~~l_1=l,~~~l_2=m \eqno(A.4)
$$
and express (A.3) as
$$
\bigg[(1-z^2)\frac{d^2}{dz^2} -(N-1)z\frac{d}{dz}
+l(l+N-2)-\frac{m(m+N-3)}{1-z^2}\bigg] Z^{(N)}_{l,m}(z)=0.
\eqno(A.5)
$$
When $N=2$, the operator ${\cal L}^2(N-1)={\cal L}^2(1)$ in the
last equation of (2.6) is given by
$$
{\cal L}^2(1)= \frac{\partial}{\partial \theta^2} \eqno(A.6)
$$
with $\theta$ denoting the corresponding $\theta_{N-1}$. Likewise,
the last equation of (3.10) can be written as
$$
-\frac{\partial^2}{\partial \theta^2} \Theta = l^2 \Theta
\eqno(A.7)
$$
with $l^2$ denoting the corresponding $l^2_{N-1}$; its solutions
will be designated as
$$
\Theta = Z^{(2)}_{l,0} = \cos l\theta \eqno(A.8)
$$
for the functions even in $\theta$, and
$$
\Theta = \overline{Z}^{(2)}_{l,0} = \sin l\theta \eqno(A.9)
$$
for the functions odd in $\theta$, with
$$
l=0,~1,~2,\cdots~. \eqno(A.10)
$$
For $N=3$, the eigenfunction of (A.5) are the Legendre polynomial
when $m=0$; i.e.,
$$
Z_{l,0}^{(3)}(z) = P_l(z) = (-)^l \frac{1}{2^l l!}
\frac{d^l}{dz^l}(1-z^2)^l \eqno(A.11)
$$
where $l=0,~1,~2,\cdots$, as before. The corresponding
$Z^{(N)}_{l,m}(z)$ for $m>0$ is given by the associated Legendre
function
$$
Z_{l,m}^{(3)}(z) = P_l^m(z) = (1-z^2)^{\frac{m}{2}}
\frac{d^m}{dz^m} P_l(z)  \eqno(A.12)
$$
with
$$
m\leq l. \eqno(A.13)
$$

In order to derive the functions $Z_{l,m}^{(N)}(z)$ for $N>3$, it
is useful to establish the following properties:

\noindent (i) Differentiate $m$ times the following equation for
$u(z)$
$$
(1-z^2)\frac{d^2u}{dz^2}-az\frac{du}{dz}+bu=0, \eqno(A.14)
$$
where $a$ and $b$ are constants, and denote
$$
v(z)=\frac{d^m}{dz^m}u(z). \eqno(A.15)
$$
We obtain
$$
(1-z^2)\frac{d^2v}{dz^2}-\alpha_m z\frac{dv}{dz}+\beta_m v=0
\eqno(A.16)
$$
with
$$
\alpha_m=a+2m
$$
$$
{\sf
and}~~~~~~~~~~~~~~~~~~~~~~~~~~~~~~~~~~~~~~~~~~~~~~~~~~~~~~~~~~~~~~~~~~~
~~~~~~~~~~~~~~~~~~~~~~~~~~~~~~~~~~ \eqno(A.17)
$$
$$
\beta_m=b-ma-m(m-1).
$$
(ii) Instead of (A.14), $u(z)$ now satisfies
$$
(1-z^2)\frac{d^2u}{dz^2}-az\frac{du}{dz}+bu-\frac{cu}{1-z^2}=0
\eqno(A.18)
$$
where $a$, $b$ and $c$ are again all constants. Write
$$
w(z)=(1-z^2)^{\frac{n}{2}} u(z). \eqno(A.19)
$$
we find
$$
(1-z^2)\frac{d^2w}{dz^2}-A_nz\frac{dw}{dz}+B_nw-\frac{C_nw}{1-z^2}=0
\eqno(A.20)
$$
with
\begin{eqnarray*}
A_n&=&a-2n\\
B_n&=&b+n(a-n-1)\\
{\sf and}~~~~~~~~~~~~~~~~~~~~~~~~~~~~~~~~~~~~~&&
 ~~~~~~~~~~~~~~~~~~~~~~~~~~~~~~~~~~~~~~~~~(A.21)\\
C_n&=&c+n(a-n-2);
\end{eqnarray*}
therefore,
$$
B_n-C_n=b-c+n. \eqno(A.22)
$$

We note that from (A.11) the Legendre polynomial
$P_l(z)=Z^{(3)}_{l,0}(z)$ satisfies (A.14) with
$$
a=2,~~~{\sf and}~~~b=l(l+1). \eqno(A.23)
$$
Thus, for $N={\sf odd}=2k+1$, we have
$$
Z^{(N)}_{l,0}(z)=\frac{d^{k-1}}{dz^{k-1}} P_{l+k-1}(z).
\eqno(A.24)
$$
It can also be readily verified that for $m\geq 0$,
$$
Z^{(N)}_{l,m}(z)=(1-z^2)^{\frac{m}{2}}\frac{d^m}{dz^m}
Z^{(N)}_{l,0}(z). \eqno(A.25)
$$
For $N={\sf even}=2k$, we start from $Z^{(2)}_{l,0}(z)= \cos
l\theta$ with $z=  \cos \theta$, and write
$$
Z^{(N)}_{l,0}(z)= \frac{d^{k-1}}{dz^{k-1}} Z^{(2)}_{l+k-1,0}(z);
\eqno(A.26)
$$
for $m>0$, the corresponding $Z^{(N)}_{l,m}(z)$ is given by the
same (A.25).

\newpage

\section*{\bf Appendix B}
\setcounter{section}{7} \setcounter{equation}{0}

Because of (4.1), $w>0$ and $w'<0$, and the hierarchy theorem, the
iterative solution with the boundary condition $f(\infty)=1$ is
convergent for any $g>0$. By examining a simple prototype example
in this class of problems, we shall show that for $g$ large, the
rate of convergence is similar to a power series in $g^{-1}$.

Consider the Schroedinger equation
$$
 -\frac{1}{2}  \psi'' +( V(x)-E)\psi =0
\eqno(B.1)
$$
in one space dimension with
$$
V(x)=\frac{g^2}{2}(x^2-1)^2 \eqno(B.2)
$$
and
$$
x \geq 0. \eqno(B.3)
$$
Throughout this appendix, ~'~ denotes $\frac{d}{dx}$.To simplify
the analysis, we impose the boundary conditions
$$
\psi(\infty)=0
$$
$$
{\sf
and~at~the~origin}~~~~~~~~~~~~~~~~~~~~~~~~~~~~~~~~~~~~~~~~~~~~~~
~~~~~~~~~~~~~~~~~~~~~~~~~~~~~~~~\eqno(B.4)
$$
$$\bigg(\frac{\psi'}{\psi}\bigg)_{x=0} = g-1.
$$
The trial function for the groundstate wave function is chosen to
be
$$
\phi(x)=\frac{2}{x+1}~e^{-gS_0}\eqno(B.5)
$$
with
$$
S_0(x)=\frac{1}{3}(x-1)^2(x+2); \eqno(B.6)
$$
it satisfies
$$
 -\frac{1}{2}  \phi'' +( U(x)-g)\phi =0
\eqno(B.7)
$$
with
$$
U(x)=V(x)+u(x),\eqno(B.8)
$$
$$
u(x)=\frac{1}{(1+x)^2}\eqno(B.9)
$$
and the same boundary conditions $\phi(\infty)=0$ and
$$
\bigg(\frac{\phi'}{\phi}\bigg)_{x=0} = g-1.\eqno(B.10)
$$
Rewrite (B.1) as
$$
 -\frac{1}{2}  \psi'' +( U(x)-g)\psi =(u(x)-{\cal E})\psi
\eqno(B.11)
$$
with
$$
E=g-{\cal E}.\eqno(B.12)
$$
Introducing
$$
f \equiv \frac{\psi}{\phi}, \eqno(B.13)
$$
we find
$$
 -\frac{1}{2} (\phi^2 f')'= ( u(x)-{\cal E})\phi^2 f;
\eqno(B.14)
$$
at $x=0$
$$
f'(0)=0.\eqno(B.15)
$$
To fix the relative normalization factor between $\psi$ and
$\phi$, we impose at $x=\infty$,
$$
f(\infty)=1.\eqno(B.16)
$$

The groundstate wave function $\psi(x)$ of the Schroedinger
equation will be solved by introducing the iterative sequences
$$
\psi_1(x),~\psi_2(x),\cdots, \psi_n(x),\cdots
$$
$$
{\sf and}~~~~~~~~~~~~~~~~~~~~~~~~~~~~~~~~~~~~~~~~~~~~~~~~~~~~~~~
~~~~~~~~~~~~~~~~~~~~~~~~~~ \eqno(B.17)
$$
$$
{\cal E}_1,~{\cal E}_2,\cdots, {\cal E}_n,\cdots
$$
with
$$
 -\frac{1}{2}  \psi''_n +( U(x)-g)\psi_n =(u(x)-{\cal
 E}_n)\psi_{n-1}
\eqno(B.18)
$$
and
$$
\psi_0=\phi. \eqno(B.19)
$$
Define
$$
f_n(x)=\frac{\psi_n(x)}{\phi(x)}~. \eqno(B.20)
$$
From the above equations, it follows that
$$
 -\frac{1}{2} (\phi^2 f'_n)'=( u-{\cal E}_n)\phi^2 f_{n-1};
 \eqno(B.21)
$$
furthermore, similar to (B.15)-(B.16),
$$
f_n'(0)=0 \eqno(B.22)
$$
and
$$
f_n(\infty)=1.\eqno(B.23)
$$
From (B.21)-(B.22), we have
$$
\int\limits_0^\infty (u(x)-{\cal E}_n)\phi^2(x)f_{n-1}(x)dx=0.
\eqno(B.24)
$$

For any function $F(x)$, define
$$
[F]=\int\limits_0^\infty F(x)\phi^2(x)dx. \eqno(B.25)
$$
From (B.24), we have
$$
{\cal E}_n = \frac{[uf_{n-1}]}{[f_{n-1}]}~. \eqno(B.26)
$$
It is convenient to regard (B.21) as an electrostatic analog
problem with $-\frac{1}{2}f'_n$ as the electrostatic field,
$$
\kappa = \phi^2 \eqno(B.27)
$$
as the dielectric constant,
$$
D_n= -\frac{1}{2} \kappa f'_n \eqno(B.28)
$$
the displacement field and
$$
\sigma _n=(u-{\cal E}_n) \phi^2 f_{n-1}\eqno(B.29)
$$
the electrostatic charge density. The electrostatic field equation
is
$$
D'_n=\sigma_n,\eqno(B.30)
$$
and (B.24) gives the condition for zero total charge,
$$
\int\limits_0^\infty \sigma_n(x)dx=0.\eqno(B.31)
$$
Consequently,
$$
D_n(0)=0.\eqno(B.32)
$$
At $x=\infty$, because $\phi(\infty)=0$, we also have
$$
D_n(\infty)=0.\eqno(B.33)
$$
Furthermore, $D_n(x)$ is positive and therefore
$$
f'_n(x)<0. \eqno(B.34)
$$
From (B.22), (B.28) and (B.30), we find
$$
D_n(x)=\int\limits_0^x \sigma_n(z)dz = -\int\limits_x^\infty
\sigma_n(z)dz \eqno(B.35)
$$
and, on account of (B.23),
$$
f_n(x)=1-2 \int\limits_x^\infty \phi^{-2}(y)dy
\int\limits_y^\infty \sigma_n(z)dz \eqno(B.36)
$$
which is equivalent to
$$
f_n(x)=f_n(0)-2 \int\limits_0^x \phi^{-2}(y)dy \int\limits_0^y
\sigma_n(z)dz.
$$

Because $u(x)$ satisfies $u(x)>0$ and $u'(x)<0$, the hierarchy
theorem applies. Therefore
$$
{\cal E}_{n-1}<{\cal E}_n\eqno(B.37)
$$
$$
\bigg(\frac{f_n}{f_{n-1}}\bigg)'<0\eqno(B.38)
$$
and
$$
\bigg(\frac{f'_n}{f'_{n-1}}\bigg)'<0. \eqno(B.39)
$$
Define
$$
g_n(x) \equiv f_n(x)-f_{n-1}(x);\eqno(B.40)
$$
for $n\geq 2$
$$
e_n \equiv {\cal E}_n-{\cal E}_{n-1}, \eqno(B.41)
$$
and for $n=1$,
$$
e_1 \equiv {\cal E}_1 -\frac{1}{4}. \eqno(B.42)
$$
For clarity, we will present our analysis in the form of several
simple theorems.\\

\noindent \underline{Theorem 1}
$$
\bigg(\frac{g_n}{f_{n-1}}\bigg)'<0,~~~\bigg(\frac{g'_n}{f'_{n-1}}\bigg)'<0,
\eqno(B.43)
$$
and
$$
g'_n<0 \eqno(B.44)
$$

\noindent \underline{Proof}~~~ From (B.40),
$$
\frac{g_n}{f_{n-1}}=\frac{f_n}{f_{n-1}}-1~~~{\sf
and}~~~\frac{g'_n}{f'_{n-1}}=\frac{f'_n}{f'_{n-1}}-1. \eqno(B.45)
$$
Therefore, (B.38) and (B.39) lead to (B.43) and therefore
$$
\bigg(\frac{g_n}{f_{n-1}}\bigg)'=\frac{g'_n}{f_{n-1}}-\frac{g_n
f'_{n-1}}{f^2_{n-1}}<0.
$$
Because $f'_{n-1}<0$ and $g_n$,~$f_{n-1}$ both positive, it
follows then
$$
g'_n<\frac{g_nf'_{n-1}}{f_{n-1}}<0. \eqno(B.46)
$$

\noindent \underline{Theorem 2}
$$
e_n>0.\eqno(B.47)
$$

\noindent \underline{Proof}~~~ For $n\geq 2$, (B.37) gives
$e_n>0$. For $n=1$, (B.26) and $f_0=1$ lead to
$$
{\cal E}_1=\frac{[u]}{[1]}.\eqno(B.48)
$$
As we shall prove
$$
[e_1]=({\cal
E}_1-\frac{1}{4})[1]=[u-\frac{1}{4}]>\frac{1}{3}e^{-\frac{4}{3}g}.\eqno(B.49)
$$

By using (B.9), we obtain
$$
u-\frac{1}{4}=\frac{3+x}{4(1+x)^2}(1-x).\eqno(B.50)
$$
Introduce
$$
\xi(x)=e^{-2gS_0}=\frac{\xi'(x)}{2g(1-x^2)}.\eqno(B.51)
$$
As $x$ varies from $0$ to $\infty$, $\xi$ follows a path ${\cal
P}$, starting from $\xi=e^{-\frac{4}{3}g}$, increasing to $\xi=1$
when $x=1$, and then decreasing  to $\xi=0$. Thus,
$$
[u-\frac{1}{4}]=\int\limits_0^\infty
\phi^2(u-\frac{1}{4})dx=\int\limits_{\cal P}
\frac{3+x}{2g(1+x)^5}d\xi. \eqno(B.52)
$$
Divide the positive $x$-axis into three sections:
$$
{\rm 1.}~~~~~~~~~~~~~~~~~~~~~~~~~x~~~{\sf from}~~~0~~{\sf
to}~~1;~~~~~~~~~~~~~~~~~~~~~~~~~~~~~~~~\eqno(B.53)
$$
correspondingly $\xi$ from $e^{-\frac{4}{3}g}$ to $1$, and $S_0$
from $\frac{2}{3}$ to $0$.

$$
{\rm 2.}~~~~~~~~~~~~~~~~~~~~~~~x~~~{\sf from}~~~1~~{\sf
to}~~\sqrt{3};~~~~~~~~~~~~~~~~~~~~~~~~~~~~~~~\eqno(B.54)
$$
$\xi$ from $1$ to $e^{-\frac{4}{3}g}$ and $S_0$ from $0$ to
$\frac{2}{3}$.

$$
{\rm 3.}~~~~~~~~~~~~~~~~~~~~~~~x~~~{\sf from}~~~\sqrt{3}~~{\sf
to}~~\infty;~~~~~~~~~~~~~~~~~~~~~~~~~~~~~~\eqno(B.55)
$$
$\xi$ from $e^{-\frac{4}{3}g}$ to $0$ and $S_0$ from $\frac{2}{3}$
to $\infty$.\\

Define
$$
F(x)=\frac{3+x}{(1+x)^5}.\eqno(B.56)
$$
Eq.(B.52) can be written as
$$
[u-\frac{1}{4}]={\rm I}(g)-{\rm II}(g)-{\rm III}(g)\eqno(B.57)
$$

\newpage

where
$$
{\rm I}(g)=\int\limits_0^{\frac{2}{3}}F(x_-(S_0))e^{-2gS_0}dS_0>0
$$
$$
{\rm II}(g)=\int\limits_0^{\frac{2}{3}}F(x_+(S_0))e^{-2gS_0}dS_0>0
\eqno(B.58)
$$
and
$$
{\rm III}(g)=\int\limits_{\frac{2}{3}}^\infty
F(x_+(S_0))e^{-2gS_0}dS_0>0,
$$
with $x_-(S_0)$ and $x_+(S_0)$ given below.

In accordance with (B.6), when $x$ varies from $-2$ to $2$, $S_0$
ranges from $0$ to $\frac{4}{3}$. In this range, for each $S_0$
there are three real roots of $x$ that satisfy
$$
x^3-3x=3S_0-2.\eqno(B.59)
$$
Let
$$
\cos \theta \equiv 1-\frac{3}{2}S_0.\eqno(B.60)
$$
With $\theta=0$ at $S_0=0$ and $x=1$. For each $S_0$ within $0$
and $\frac{4}{3}$ designate these three roots as
$$
x_+(S_0)=2\cos \alpha_+(S_0),~ x_-(S_0)=2\cos \alpha_-(S_0)
$$
$$
{\sf and}~~~~~~~~~~~~~~~~~~~~~~~~~~~x_0(S_0)=2\cos
\alpha_0(S_0),~~~~~~~~~~~~~~~~~~~~~~~~~~~~~~~\eqno(B.61)
$$
with
$$
\alpha_+(S_0)=-60^0 + \frac{\theta}{3}
$$
$$
\alpha_-(S_0)=60^0 + \frac{\theta}{3}\eqno(B.62)
$$
$$
\alpha_0(S_0)=180^0 + \frac{\theta}{3}.
$$
The $x$ referred to in (B.53) and (B.54)-(B.55) are respectively
the above $x_-(S_0)$ and $x_+(S_0)$ together with its analytical
extension to $x_+>2$.

\newpage

When $g=0$, by using
$$
-S_0'=1-x^2
$$
$$
{\sf and}~~~~~~~~~~~~~~~~~~~~~~~~~~~~~~~~~~~~~~~~~~~~~~~~~~
~~~~~~~~~~~~~~~~~~~~~~~~~~~~~~~~~~~~\eqno(B.63)
$$
$$
(3+x)(1-x^2)=(1+x)\bigg(4-(1+x)^2\bigg),
$$
we have from (B.52) and (B.57)
$$[u-\frac{1}{4}]_{g=0}=\int\limits_0^\infty
\frac{1}{(1+x)^4}\bigg(4-(1+x)^2\bigg)dx
$$
$$
~~~~~~~~~={\rm I}(0)- {\rm II}(0)- {\rm III}(0)=\frac{1}{3}.
\eqno(B.64)
$$
Take the expression for ${\rm III}(g)$ in (B.58). Since in its
integrand $e^{-2gS_0}<e^{-\frac{4}{3}g}$, we have
$$
{\rm III}(g)<e^{-\frac{4}{3}g}{\rm III}(0).\eqno(B.65)
$$
Next consider
$$
{\rm I}(g)- {\rm II}(g)=\int\limits_0^{\frac{2}{3}}
\bigg(F(x_-(S_0))-F(x_+(S_0))\bigg)e^{-2gS_0}dS_0 \eqno(B.66)
$$
in which, since $x_+(S_0)>x_-(S_0)$,
$$
F(x_-(S_0))-F(x_+(S_0))>0. \eqno(B.67)
$$
In addition, because in its integrand
$$
e^{-2gS_0}>e^{-\frac{4}{3}g},\eqno(B.68)
$$
we find
$$
{\rm I}(g)- {\rm II}(g)>e^{-\frac{4}{3}g} \bigg({\rm I}(0)- {\rm
II}(0)\bigg). \eqno(B.69)
$$
Combining with (B.64)-(B.65), we derive
$$
{\rm I}(g)-{\rm II}(g)-{\rm III}(g)>e^{-\frac{4}{3}g} \bigg({\rm
I}(0)- {\rm II}(0)- {\rm
III}(0)\bigg)>\frac{1}{3}e^{-\frac{4}{3}g} \eqno(B.70)
$$
which leads to (B.49) and $e_1>0$, and thereby completes the proof
of\\ Theorem 2.

\noindent \underline{Theorem 3}
$$
e_1=\frac{[u-\frac{1}{4}]}{[1]}
$$
and for $n \geq 2$
$$
e_n=\frac{[(u-{\cal
E}_{n-1})g_{n-1}]}{[f_{n-1}]}<\frac{[(u-\frac{1}{4})g_{n-1}]}{[1]}.
\eqno(B.71)
$$
\underline{Proof}~~~ The first equation for $n=1$ follows from
(B.42). For $n \geq 2$,
$$
e_n={\cal E}_n-{\cal E}_{n-1}=\frac{[(u-{\cal
E}_{n-1})f_{n-1}]}{[f_{n-1}]}. \eqno(B.72)
$$
In accordance with (B.40) and (B.26),
$$
f_{n-1}=f_{n-2}+g_{n-1} \eqno(B.73)
$$
and
$$
[(u-{\cal E}_{n-1})f_{n-2}]=0.\eqno(B.74)
$$
Thus, the equality in (B.71) follows. Since for all $n$,
$$
{\cal E}_n \geq {\cal E}_1 >\frac{1}{4}, \eqno(B.75)
$$
we establish also the inequality in (B.71).\\

\noindent \underline{Theorem 4}
$$
\frac{[(u-\frac{1}{4})g_n]}{[1]} < \frac{1}{8g}(14g_n(0)+3G_n),
\eqno(B.76)
$$
where
$$
G_n \equiv {\sf maximum~~of}~~-g'_n(x).\eqno(B.77)
$$
\underline{Proof} ~~~As in (B.52), we write
$$
[(u-\frac{1}{4})g_n]=\int\limits_0^\infty
\frac{3+x}{2g(1+x)^5}g_n(x) \frac{d}{dx}(e^{-2gS_0})dx
$$
$$
~~~~~~~~=-\frac{3}{2g}g_n(0)e^{-\frac{4}{3}g}
+\frac{1}{8g}\int\limits_0^\infty \phi^2({\rm A}g_n-{\rm B}g'_n)dx
\eqno(B.78)
$$
with
$$
{\rm A}(x)= \frac{3+x}{(1+x)^3}
\bigg(\frac{5}{1+x}-\frac{1}{3+x}\bigg)= \frac{2(7+2x)}{(1+x)^4}
\leq 14 \eqno(B.79)
$$

\newpage

and
$$
{\rm B}(x)= \frac{3+x}{(1+x)^3} \leq 3. \eqno(B.80)
$$
Since the first term on the right hand side of (B.78) is negative,
Theorem 4 is proved.\\

\noindent \underline{Theorem 5}~~~ For $0<x \leq 1$,
$$
-g'_n(x) < \frac{3}{g}~ g_{n-1}(0). \eqno(B.81)
$$
\underline{Proof}~~~ Integrating (B.21) from $0$ to $x$, and using
(B.22), we obtain
$$
-\frac{1}{2}\phi^2(x)f'_n(x)=\int\limits_0^x \phi^2(z)(u(z)-{\cal
E}_n)f_{n-1}(z)dz
$$
$$
{\sf and}~~~~~~~~~~~~~~~~~~~~~~~~~~~~~~~~~~~~~~~~~~~~~~~
~~~~~~~~~~~~~~~~~~~~~~~~~~~~~~~~~~~~\eqno(B.82)
$$
$$
-\frac{1}{2}\phi^2(x)f'_{n-1}(x)=\int\limits_0^x
\phi^2(z)(u(z)-{\cal E}_{n-1})f_{n-2}(z)dz.
$$
The difference between these two equations gives
$$
-\frac{1}{2}\phi^2(x)g'_n(x)=\int\limits_0^x
\phi^2(z)\bigg\{(u(z)-{\cal E}_n)g_{n-1}(z)-e_nf_{n-2}(z)\bigg\}dz
$$
$$
~~~~~~~~~~< \int\limits_0^x \phi^2(z)(u(z)-{\cal
E}_n)g_{n-1}(z)dz,\eqno(B.83)
$$
since $-e_nf_{n-2}(z)<0$. From (B.44) and (B.49) it follows then
$$
g_{n-1}(z)<g_{n-1}(0),\eqno(B.84)
$$
$$
{\cal E}_n \geq {\cal E}_1 > \frac{1}{4} \eqno(B.85)
$$
and (B.83) becomes
$$
-\frac{1}{2}\phi^2(x)g'_n(x)<\int\limits_0^x
\phi^2(z)\bigg(u(z)-\frac{1}{4}\bigg)g_{n-1}(z)dz
$$
$$
~~~~~~~~~~~<g_{n-1}(0)\int\limits_0^x
\phi^2(z)\bigg(u(z)-\frac{1}{4}\bigg)dz
$$
$$
~~~~~~~~~~~=g_{n-1}(0)\int\limits_0^x
\frac{3+x}{2g(1+x)^5}(e^{-2gS_0})'dx.\eqno(B.86)
$$
Because
$$
\bigg(\ln \frac{3+x}{(1+x)^5}\bigg)'=-\frac{14+4x}{(3+x)(1+x)}<0,
$$
(B.86) leads to
$$
-\frac{1}{2}\phi^2(x)g'_n(x)<\frac{3g_{n-1}(0)}{2g}\bigg(e^{-2gS_0(x)}-e^{-\frac{4}{3}g}\bigg)
$$
$$
~~~~~~~~~~~~~~<\frac{3g_{n-1}(0)}{2g} e^{-2gS_0(x)}.
$$
By using (B.5), we derive
$$
-g'_n(x)<\frac{3}{g}~\frac{(1+x)^2}{4} g_{n-1}(0) \eqno(B.87)
$$
which for $0<x<1$ leads to Theorem 5.\\

\noindent \underline{Lemma} ~~~When $n=1$ and $x> 1$,
$$
\bigg(-(1+x)^2f'_1\bigg)'<0 \eqno(B.88)
$$
provided
$$
g> 2. \eqno(B.89)
$$
The prove of the lemma is given at the end of this Appendix. We
will now proceed assuming its validity.\\

\noindent \underline{Theorem 6} ~~~For $x>1$ and $g> 2$,
$$
\bigg(-(1+x)^2f'_n\bigg)'<0 \eqno(B.90)
$$
and
$$
\bigg(-(1+x)^2g'_n\bigg)'<0. \eqno(B.91)
$$
\underline{Proof}~~~ Since
$$
-(1+x)^2f'_n=\frac{f'_{n}}{f'_{n-1}}\cdot\frac{f'_{n-1}}{f'_{n-2}}\cdots
\frac{f'_{2}}{f'_{1}}\bigg(-(1+x)^2f'_1\bigg) ~~~~\eqno(B.92)
$$
and
$$
-(1+x)^2g'_n=\frac{g'_{n}}{f'_{n-1}}\cdot\frac{f'_{n-1}}{f'_{n-2}}\cdots
\frac{f'_{2}}{f'_{1}}\bigg(-(1+x)^2f'_1\bigg). \eqno(B.93)
$$
Using (B.39), (B.43) and the lemma, we see that the derivatives of
(B.92) and (B.93) are negative.\\

\noindent \underline{Theorem 7} ~~~For $g> 2$ and $x> 1$,
$$
G_n={\sf
max}~\bigg(-g'_n(x)\bigg)<\frac{3}{g}~g_{n-1}(0).\eqno(B.94)
$$
\underline{Proof} ~~~For $x>1$,
$$
\bigg(-g'_n(x)\bigg)'<0\eqno(B.95)
$$
on account of (B.91); therefore
$$
-g_n'(x)<-g'_n(1).\eqno(B.96)
$$
Together with Theorem 5, (B.94) is proved.\\

\noindent \underline{Theorem 8} ~~~For $g> 2$ and for all $x>0$,
$$
g_n(x)<g_n(0)<\frac{9}{g}~g_{n-1}(0)<\cdots<\bigg(\frac{9}{g}\bigg)^n
\eqno(B.97)
$$
\underline{Proof} ~~~Since
$$
g_n(0)=g_n(1)-\int\limits_0^1 g'_n(x)dx<g_n(1)+G_n \eqno(B.98)
$$
and
$$
g_n(1)=-\int\limits_1^\infty g'_n(x)dx=\int\limits_1^\infty
[-(1+x)^2 g'_n]\frac{dx}{(1+x)^2}~.\eqno(B.99)
$$
On account of (B.91), for $x>1$
$$
-(1+x)^2g'_n(x)<-4g'_n(1); \eqno(B.100)
$$
therefore (B.99) yields
$$
g_n(1)<-4g'_n(1)\int\limits_1^\infty \frac{dx}{(1+x)^2} =
-2g'_n(1)<2G_n. \eqno(B.101)
$$
Combining this result with (B.98), we derive
$$
g_n(0)<3G_n<\frac{9}{g}~g_{n-1}(0),\eqno(B.102)
$$
on account of (B.94).\\

\noindent \underline{Theorem 9} ~~~For $g> 2$,
$$
e_n<\frac{5}{24}~\bigg(\frac{9}{g}\bigg)^n\eqno(B.103)
$$
\underline{Proof} ~~~Using (B.76), (B.94) and (B.97), we derive
$$
\frac{[(u-\frac{1}{4})g_{n-1}]}{[1]}<\frac{1}{8g}(14g_{n-1}(0)+3G_{n-1})
$$
$$
~~~~~~~~~~<\frac{1}{8g}\bigg(14\bigg(\frac{9}{g}\bigg)^{n-1}
+\bigg(\frac{9}{g}\bigg)^{n-1}\bigg)=\frac{15}{8g}\bigg(\frac{9}{g}\bigg)^{n-1}
$$
$$
~~~~~~~~~=\frac{5}{24}\bigg(\frac{9}{g}\bigg)^n.
$$
Substituting this result into (B.71), we complete the proof of
Theorem 9.

We will now turn to the proof of the lemma. For $n=1$, (B.29) and
(B.36) lead to
$$
f'_1(x)= 2  \phi^{-2}(x) \int\limits_x^\infty (u(z)-{\cal
E}_1)\phi^2(z)dz. \eqno(B.104)
$$
Define
$$
F(x) \equiv -\frac{1}{2}(1+x)^2f'_1(x)\eqno(B.105)
$$
$$
\xi(x) \equiv \frac{\phi^2(x)}{4(1+x)^2}=
\frac{1}{(1+x)^4}e^{-2gS_0(x)}\eqno(B.106)
$$
and
$$
\eta(x)=-\frac{1}{4}\int\limits_x^\infty (u(z)-{\cal
E}_1)\phi^2(z)dz. \eqno(B.107)
$$
We find
$$
F=\eta/\xi,\eqno(B.108)
$$
with
\begin{eqnarray*}
\eta'&=&-\xi p,~~~~~~~~~~~~~~~~~~~~~~~~~~~~~~~~~~~~~~~~~~(B.109)\\
\xi'&=& -\xi q,~~~~~~~~~~~~~~~~~~~~~~~~~~~~~~~~~~~~~~~~~~(B.110)\\
p&=&(1+x)^2{\cal E}_1-1~~~~~~~~~~~~~~~~~~~~~~~~~~~~~~(B.111)\\
{\sf and}~~~~~~~~~~~~~~~~~~~~~~~~~~q&=&\frac{4}{1+x}
+2g(x^2-1).~~~~~~~~~~~~~~~~~~~~~~~(B.112)
\end{eqnarray*}
Since ${\cal E}_1>\frac{1}{4}$ in accordance with (B.49), for
$x>1$, we have
$$
u(x)=\frac{1}{(x+1)^2}<{\cal E}_1.\eqno(B.113)
$$
Therefore
$$
\eta(x)>0;
$$
in addition,
$$
p(x)>0~~~{\sf and}~~~q(x)>0.\eqno(B.114)
$$
Regarding $\eta=\eta(\xi)$, we define
$$
L(\xi) \equiv \xi\frac{d\eta}{d\xi}-\eta.\eqno(B.115)
$$
Its derivative is
$$
\frac{dL}{d\xi}=\xi\frac{d^2\eta}{d\xi^2}=\frac{pq'-qp'}{q^3}~.\eqno(B.116)
$$
Likewise, by differentiating (B.108), we derive
$$
\xi^2\frac{dF}{d\xi}=\xi\bigg(\frac{d\eta}{d\xi}-\frac{\eta}{\xi}\bigg)=L.
\eqno(B.117)
$$

At $x=\infty$, (B.106) and (B.107) yield
$$
\xi(\infty)=0~~~{\sf and}~~~\eta(\infty)=0;\eqno(B.118)
$$
their ratio is
$$
F(\infty)=\frac{\eta(\infty)}{\xi(\infty)}=\frac{\eta'(\infty)}{\xi'(\infty)}=
\frac{p(\infty)}{q(\infty)}=\frac{{\cal E}_1}{2g}>0.\eqno(B.119)
$$
From (B.115), we see that at $x=\infty$
$$
\bigg(L(\xi)\bigg)_{x=\infty}=L(0)=0.\eqno(B.120)
$$
As $x$ decreases, since $\xi'$ and $\eta'$ both are negative,
$\xi(x)$ and $\eta(x)$ increase and are both positive. Next, we
shall prove that for $x>1$ and $g > 2$
$$
\lambda(x) \equiv \frac{1}{4}(pq'-qp')>0;\eqno(B.121)
$$
i.e.,
$$
\frac{dL}{d\xi}=\frac{4}{q^3}\lambda>0,\eqno(B.122)
$$
$$
L>0\eqno(B.123)
$$
and therefore, because of (B.117),
$$
\frac{dF}{d\xi}>0.\eqno(B.124)
$$
In turn, because of (B.110), for $x>1$
$$
F'=-\xi q\frac{dF}{d\xi}<0\eqno(B.125)
$$
and that leads to, on account of (B.105)
$$
-\bigg((1+x)^2f'_1\bigg)'<0
$$
which is the lemma (B.88)-(B.89).

To establish (B.121), we use the identity
$$
\int\limits_0^\infty
\phi^2(x)\bigg(\frac{1}{(1+x)^n}-\frac{1}{2^n}\bigg)dx=
-\frac{2^n-1}{2^{n-1}g}e^{-\frac{4}{3}g}
$$
$$
~~~~~~~~+ \frac{1}{4g}\int\limits_0^\infty
\phi^2(x)\bigg(\frac{n+3}{(1+x)^{n+2}}+\frac{n+2}{2(1+x)^{n+1}}
$$
$$
~~~~~~~~+\frac{n+1}{2^2(1+x)^{n}}+\cdots
+\frac{4}{2^{n-1}(1+x)^3}\bigg)dx. \eqno(B.126)
$$
It follows then
$$
\int\limits_0^\infty
\phi^2(x)\bigg(\frac{1}{(1+x)^2}-\frac{1}{4}\bigg)dx=
-\frac{3}{2g}e^{-\frac{4}{3}g}
$$
$$
~~~~~~~~~~+ \frac{1}{4g}\int\limits_0^\infty
\phi^2(x)\bigg(\frac{5}{(1+x)^4}+\frac{4}{2(1+x)^3}\bigg)dx,
\eqno(B.127)
$$
$$
\int\limits_0^\infty
\phi^2(x)\bigg(\frac{1}{(1+x)^2}-\frac{1}{4}-\frac{9}{2^6g}\bigg)dx=
-\bigg(\frac{3}{2g}+\frac{103}{2^5g^2}\bigg)e^{-\frac{4}{3}g}
$$
$$
~~~~~~~+ \frac{1}{2^4g^2}\int\limits_0^\infty
\phi^2(x)\bigg(\frac{35}{(1+x)^6}+\frac{54}{2(1+x)^5}+\frac{45}{2^2(1+x)^4}
+\frac{36}{2^3(1+x)^3}\bigg)dx. \eqno(B.128)
$$
and
$$
\int\limits_0^\infty
\phi^2(x)\bigg(\frac{1}{(1+x)^2}-\frac{1}{4}-\frac{9}{2^6g}
-\frac{85}{2^9g^2}\bigg)dx =[1]\frac{\delta}{g^3}\eqno(B.129)
$$
where
$$
[1]\frac{\delta}{g^3}= -\bigg(\frac{3}{2g}+\frac{103}{2^5g^2}
+\frac{2403}{2^8g^3}\bigg)e^{-\frac{4}{3}g}
$$
$$
~~~~~~~~+\frac{1}{2^6g^3} \int\limits_0^\infty
\phi^2(x)\bigg(\frac{315}{(1+x)^8}+\frac{712}{2(1+x)^7}+\frac{938}{2^2(1+x)^6}
+\frac{1020}{2^3(1+x)^5}
$$
$$
~~~~~~~~+\frac{850}{2^4(1+x)^4}+\frac{680}{2^5(1+x)^3}\bigg)dx.
\eqno(B.130)
$$
Next, using (B.48), we can write the above expression (B.129) as
$$
{\cal
E}_1=\frac{1}{4}+\frac{9}{2^6g}+\frac{\gamma}{g^2}\eqno(B.131)
$$
where
$$
\gamma=\frac{85}{2^9}+\frac{\delta}{g}\eqno(B.132)
$$
with $\delta$ defined by (B.130). From (B.111)-(B.112), we find
that $\lambda(x)$ defined by (B.121) is given by
$$
\lambda(x)=g{\cal E}_1(x+1)^2-gx+\frac{1}{(1+x)^2}-3{\cal
E}_1.\eqno(B.133)
$$
Its derivatives are
$$
\lambda'(x)=2g{\cal E}_1(x+1)-g-\frac{2}{(1+x)^3}\eqno(B.134)
$$
and
$$
\lambda''(x)=2g{\cal E}_1+\frac{6}{(1+x)^4}>0.\eqno(B.135)
$$
At $x=1$,
$$
\lambda'(1)=4g{\cal E}_1-g-\frac{1}{4}.\eqno(B.136)
$$
Using (B.131), we find
$$
\lambda'(1)=\frac{5}{16}+\frac{4\gamma}{g}.\eqno(B.137)
$$
Neglecting $O(e^{-\frac{4}{3}g})$, $\delta$ is positive; therefore
$\gamma>0$ and $\lambda'(1)>0$. Because $\lambda''(x)>0$,
$\lambda'(x)>0$ for all $x>1$, and the minimum of $\lambda(x)$ is
at $x=1$ with
$$
\lambda(1)=\frac{1}{16}+\frac{4}{g}\bigg(\gamma-\frac{27}{256}\bigg)
-\frac{3\gamma}{g^2}.\eqno(B.138)
$$
In order that $\lambda(1)>0$, we require
$$
64\gamma >g\bigg(\frac{27-4g}{4g-3}\bigg)~.\eqno(B.139)
$$
Assuming $g$ not too small so that we can neglect $\delta/g$ in
(B.132); hence
$$
\gamma \cong \frac{85}{2^9}\eqno(B.140)
$$
and
$$
64\gamma \cong \frac{85}{8}=10.625.\eqno(B.141)
$$
The righthand side of (B.139) is
$$
g\bigg(\frac{27-4g}{4g-3}\bigg)=\frac{21}{2}~~~{\sf
when}~~~g=1.5\eqno(B.142)
$$
consistent with the inequality (B.139). If in (B.132), we take
$\delta/g$ into account, then a sufficient condition for
$\lambda(1)>0$ and therefore $\lambda(x)>0$ for $x\geq 1$ is
$$
g>2. \eqno(B.143)
$$
This, together with (B.122)-(B.125) complete the proof of the
lemma.

\newpage

\vspace{1cm}

Table 1\\

\begin{tabular}{|c|c|c|c|c|c|}
  \hline
  $g=3$ &  &  &  &$g^2_{max}$&$g^2_{min}$   \\
  \hline
  $k$ & $a$ &$a_{max}$ &$a_{min}$& $(\frac{k}{a}+1)(\frac{k}{a}+\frac{1}{a}+1)$ &$(\frac{k}{a}+1)^2$\\
  \hline
 0.5 &  0.4 &.46&.25& 10.69 & 5.06\\
  \hline
  1 & 0.6 &.72&.5& 11.56 & 7.13 \\
  \hline
  1.5  & 0.8 &.98&.75& 11.86 & 8.29 \\
  \hline
  2 &  1.2 &1.23&1.0 & 9.33 & 7.13 \\
  \hline
  2.5 &  1.3 &1.49&1.25 & 10.79 & 8.52  \\
  \hline
  3 &  1.6 &1.74&1.5 & 10.06 & 8.29   \\
  \hline
  3.5 &1.8 & 1.97&1.75& 10.31 & 8.64\\
  \hline
  4 &  2.1 &2.24&2.0& 9.82 & 8.41  \\
  \hline
\end{tabular}

\vspace{1cm}

{\large \bf Figure caption}\\

Fig. 1. Iterative radial wave functions $\psi_n(r)$ and their
corresponding energies $E_n$ for both the boundary conditions
$f_n(0)=1$ (upper curves) and $f_n(\infty)=1$ (lower curves). The
parameters are $g=3$, $k=2$ and $a=1.2$.

Fig. 2. Final ${\cal R}(r)={\cal R}_{N,l}(r)$ and $E=E_{N,l}$ that
satisfy (2.11) for $g=3$, $l=0$ and $N=3,~4,~5,~6$. The overall
normalization factor for these curves are determined by requiring
$f(\infty)=1.1,~1.1~,1.0~{\sf and}~0.7$ for the corresponding
$N=3,~4,~5~{\sf and}~6$.

Fig. 3. Final ${\cal R}(r)={\cal R}_{N,l}(r)$ and $E=E_{N,l}$ that
satisfy (2.11) for $g=3$, $N=3$ and $l=0,~1,~2,~3$. The overall
normalization factor for these curves are all determined by
requiring $f(\infty)=1$.\\

{\large \bf Table caption}\\

Table 1.  List of parameter $a$ used in Figures 1-3 for different
$k=l+\frac{1}{2}(N-1)$, but with the same $g=3$. The two last
columns are $g^2_{max}=(\frac{k}{a}+1)(\frac{k}{a}+\frac{1}{a}+1)$
and $g^2_{min}=(\frac{k}{a}+1)^2$ for different pairs $(k,~a)$
used in these figures.
\newpage

\centerline{\epsfig{file=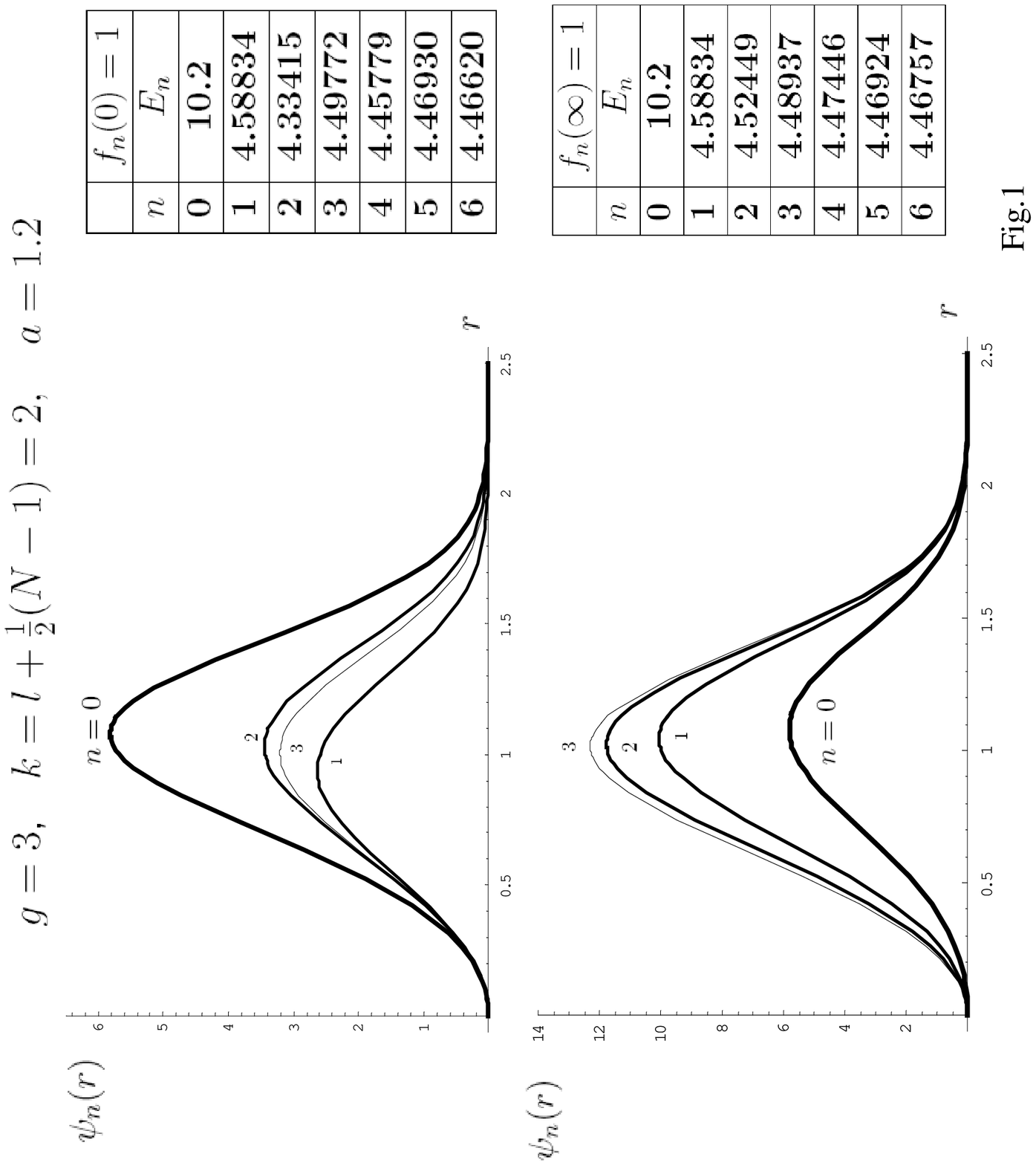,width=13cm}}
\centerline{\epsfig{file=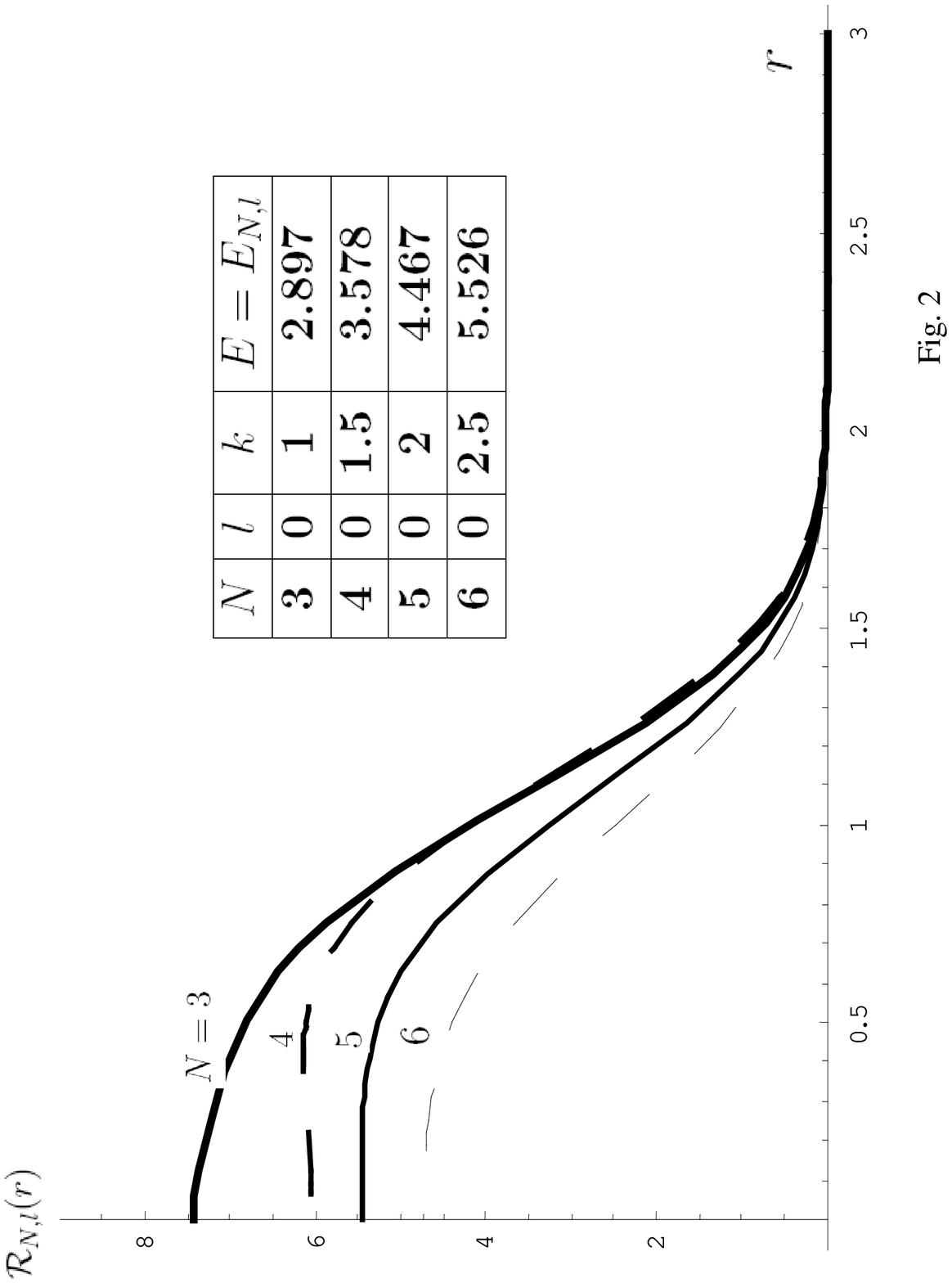,width=13cm}}
\centerline{\epsfig{file=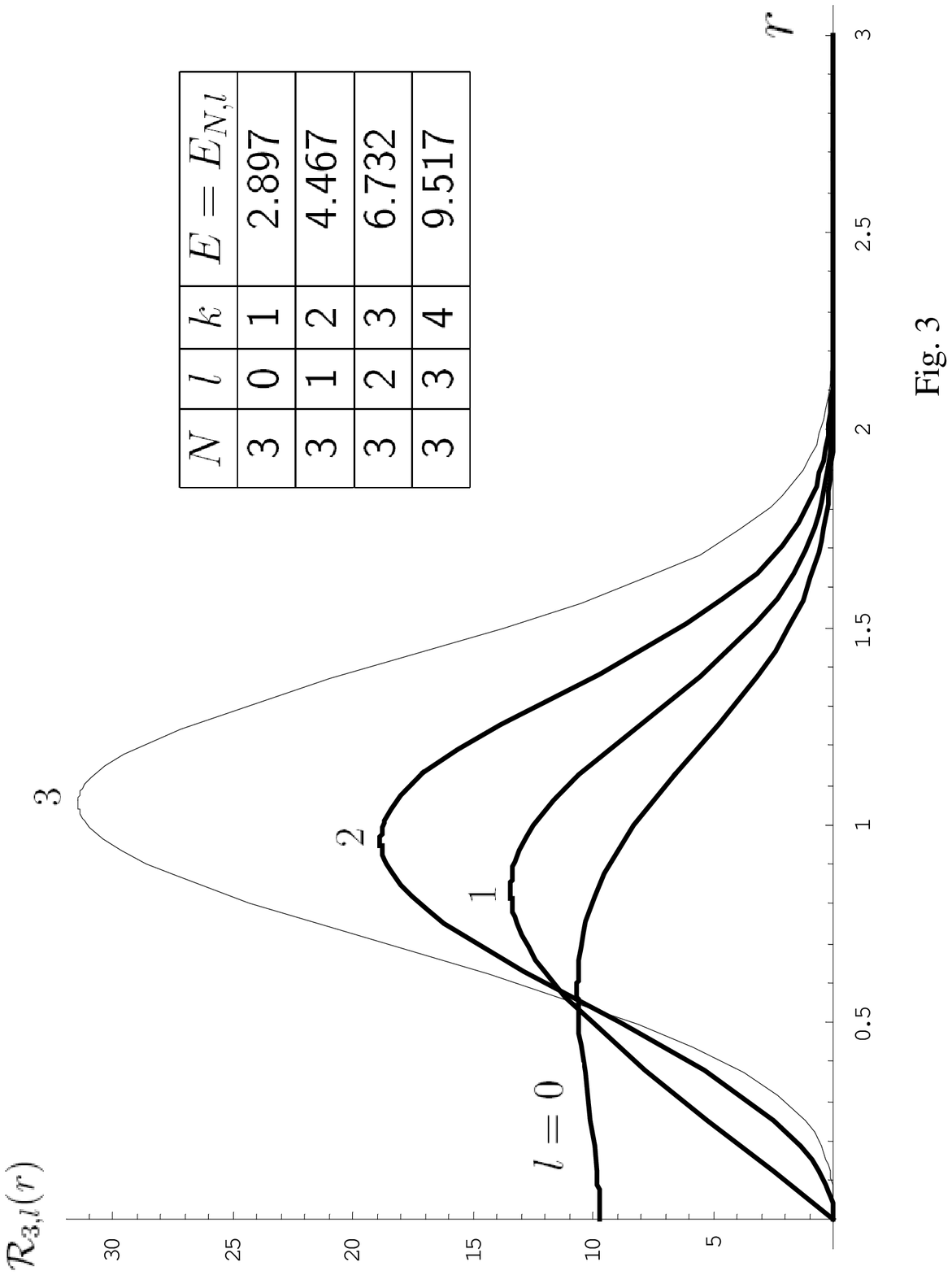,width=14cm}}

\newpage

\end{document}